\documentstyle[12pt]{article}
\begin{document}
\begin{center}

{\large\bf The HMW effect in Noncommutative Quantum Mechanics}
\vskip 1cm Jianhua Wang$^{a,c}$, Kang
Li$^{b,c}$\footnote{kangli@hztc.edu.cn}
\\\vskip 1cm

{\it\small$^a$ Department of Physics, Shaanxi University  of
Technology , Hanzhong, 723001,P.R. China}\\
$^b $Department of Physics, Hangzhou Teachers College,Hangzhou,
310036, P.R.China\\
$^c$The Abdus Salam ICTP, P.O. Box 586, 34014 Trieste, Italy
 \vskip 0.5cm
\end{center}

\begin{abstract}
\noindent The HMW effect in non-commutative quantum mechanics is
studied. By solving the Dirac equations on non-commutative (NC)
space and non-commutative phase space, we obtain topological HMW
phase on NC space and NC phase space respectively, where the
additional terms related to the space-space and momentum-momentum
non-commutativity are given explicitly.

PACS number(s): 02.40.Gh, 11.10.Nx, 03.65.-w
\end{abstract}

\section{Introduction}

The study of physics effects on non-commutative space has
attracted much attention in recent years. Because the effects of
the space non-commutativity may become significant not only in the
string scale but also at the very high  energy level (Tev and
higher energy level ). Besides the field theory,  there are many
papers devoted to the study of various aspects of quantum
mechanics on NC space with usual (commutative) time coordinate.
For example, the topological AB and AC effects on NC space and
even on NC phase space have been studied \cite{1}-\cite{6}. In
this paper we will deal with another very interesting topological
effect, HMW effect, on NC space and NC phase space respectively.
The HMW effect was firstly discussed in 1993 by He and
Meckellar\cite{7} and a year later, independently by
Wilkens\cite{9}. The HMW effect corresponds to a topological phase
related to a neutral spin-1/2 particle with non-zero electric
dipole moving in the magnetic field, and in 1998, Dowling,
Willianms and Franson point out that the HMW effect can be
partially tested using metastable hydrogen atoms\cite{10}. Just as
the AB AC effect, the HMW effect has the same importance in the
literature, and the study of the correction of the space (and
momenta) non-commutativity to the HMW effect will be meaningful.

To begin with, let's first give a brief review of some properties
of non-commutative space and non-commutative phase space. In NC
space the coordinate and momentum operators satisfy the following
commutation relations(we set $\hbar=c=1 $ in this paper )
\begin{equation}\label{eq1}
~[\hat{x}_{i},\hat{x}_{j}]=i\theta_{ij},~~~
[\hat{p}_{i},\hat{p}_{j}]=0,~~~[\hat{x}_{i},\hat{p}_{j}]=i
\delta_{ij},
\end{equation}
where $\hat{x}_i$ and $\hat{p}_i$ are the coordinate  and momentum
operators on a NC space. When a spin-$1/2$ particle moves in a
electro-magnetic field, the Dirac equation for the particle,
 usually, can be written as
 $[i\gamma_\mu\partial^\mu+S_\mu\gamma^\mu-m]\psi=0$, the $S_\mu$ here is  a
 Lorentz vector depends not only on the electro-magnetic field in which the particle
 moves but also on the electro-magnetic properties of the particle
 itself. On the NC  space, this Dirac equation becomes
\begin{equation}\label{eq2}
[i\gamma_\mu\partial^\mu+S_\mu\gamma^\mu-m]\star\psi=0,
\end{equation}
i.e. just replace normal product to a star product, then the Dirac
equation in commuting space will change into the Dirac equation in
NC space. The Moyal-Weyl (or star) product between two functions
is defined by,
\begin{equation}\label{eq3}
(f  \ast g)(x) = e^{  \frac{i}{2}
 \theta_{ij} \partial_{x_i} \partial_{x_j}
 }f(x_i)g(x_j)  = f(x)g(x)
 + \frac{i}{2}\theta_{ij} \partial_i f \partial_j
 g\big|_{x_i=x_j}+{\mathcal{ O}}(\theta^{2}),
\end{equation}
here $f(x)$ and $g(x)$ are two arbitrary functions. Other than to
solve the NC Dirac equation by using the star product, a
equivalent method will be used in this paper, that is,  we replace
the star product in Dirac equation  with usual product by shift NC
coordinates with a Bopp's shift
\begin{equation}\label{eq4}
 \hat{x}_{i}=  x_{i}-\frac{1}{2}\theta_{ij}p_{j} ,
 ~~\hat{p_i}=p_i,
\end{equation}
as well as a shift for the vector for vector $S_\mu$,
\begin{equation}\label{50}
 S_\mu\rightarrow\hat{S}_\mu= S_\mu +\frac{1}{2}
 \theta^{\alpha\beta} p_\alpha\partial_\beta S _\mu .
\end{equation}
Then the Dirac equation can be solved in the commuting space and
the non-commutative properties can be realized through the terms
related to $\theta$.

 The Bose-Einstein statistics in non-commutative quantum mechanics
 requires both space-space and  momentum-momentum non-commutativity,   the
 space in
 this case is called NC phase
space. On NC phase space, the commutation relations (\ref{eq1})
should be replaced with
\begin{equation}\label{eq5}
[\hat{x}_{i},\hat{x}_{j}]=i\theta_{ij},~~
[\hat{p}_{i},\hat{p}_{j}]=i\bar{\theta}_{ij},~~[\hat{x}_{i},\hat{p}_{j}]=i
\delta_{ij}.
\end{equation}
and the star product in Eqs. (\ref{eq2}) defines,
\begin{eqnarray}\label{eq6}
(f  \ast g)(x,p) = e^{ \frac{i}{2\alpha^2}
 \theta_{ij} \partial_i^x \partial_j^x+\frac{i}{2\alpha^2}\bar{\theta}_{ij} \partial_i^p
 \partial_j^p}
 f(x,p)g(x,p) ~~~~~~~~~~~~~~~~~~~~~~~\nonumber\\ = f(x,p)g(x,p)
 + \frac{i}{2\alpha^2}\theta_{ij} \partial_i^x f \partial_j^x g\big|_{x_i=x_j}
 + \frac{i}{2\alpha^2}\bar{\theta}_{ij} \partial_i^p f \partial_j^p g\big|_{p_i=p_j}+{\mathcal{O}}(\theta^2).
\end{eqnarray}
Where ${\mathcal{O}}(\theta^2)$ stands the second and higher order
terms of $\theta$ and $\bar{\theta}$. To replace the star product
in Dirac equation on NC phase space we need a  generalized Bopp's
shift
\begin{eqnarray}\label{eq7}
x_\mu\rightarrow \alpha x_{i}-\frac{1}{2 \alpha}\theta_{\mu\nu}p_{\nu},\nonumber\\
p_\mu\rightarrow \alpha
p_\mu+\frac{1}{2\alpha}\bar{\theta}_{\mu\nu}x_{\nu},
\end{eqnarray}
and together with a shift,
\begin{equation}\label{52}
 S_\mu\rightarrow\hat{S}_\mu= \alpha S_\mu +\frac{1}{2\alpha}
 \theta^{\alpha\beta} p_\alpha\partial_\beta S _\mu ,
\end{equation}
which is the  partner of shift in  Eq.(\ref{50}) on NC phase
space.

\section{Review of the HMW effect on 2+1 dimensional commutative space time }

In order to study the NC properties of HMW effect, a brief review
of the effect in $2+1$ dimensional commutative space time is
necessary. The Lagrange of a spin-1/2 neutral particle with
electric dipole $\mu_e$ moving in the electromagnetic field is
given by
\begin{eqnarray}\label{eq8}
L = \bar \psi i \gamma^\mu \partial_\mu \psi - m\bar \psi \psi -
i{1\over 2} \mu_e\bar \psi \sigma^{\mu\nu}\gamma_5 \psi
F_{\mu\nu}. \label{Lagrange}
\end{eqnarray}
The last term in the Lagrangian represents the HMW effect. Using
the identity  $\sigma^{\mu\nu} \gamma_5 =
(i/2)\epsilon^{\mu\nu\alpha\beta} \sigma_{\alpha\beta}$, the
Lagrangian  becomes
\begin{eqnarray}\label{eq9}
L = \bar \psi i \gamma^\mu \partial_\mu \psi - m\bar \psi \psi +
{1\over 2}\mu_e \tilde{F}_{\mu\nu} \bar \psi\sigma^{\mu\nu}\psi ,
\label{Lagrange2}
\end{eqnarray}
where $\tilde{F}$ is the Hodge star of $F$, i.e.
$\tilde{F}_{\mu\nu}=\frac{1}{2}\epsilon_{\mu\nu\alpha\beta
}F^{\alpha\beta}$.  Similar as AB, AC other topological effects,
the HMW effect is also usually studied in $2+1$ dimension, because
the particle movies in a plane.

We restrict the particle moves on a plane (say $x-y$ plane), then
the problem can be treated in $2+1$ space time. We use the
conventions  $g_{\mu\nu}={\rm diag}(1,-1,-1)$ and the
anti-symmetric tensor $\epsilon_{\mu\nu\alpha}$ with
$\epsilon_{012} = +1$. We will use $3$ four component Dirac
matrices which can describe spin up and down in the notional $z$
direction for a particle and for its anti-particle\cite{18}.  In
2+1 dimensions these Dirac matrices satisfy the following relation
\begin{equation}\label{eq11}
    \gamma^\mu \gamma^\nu = g^{\mu\nu}
-i\gamma^0\sigma^{12}\epsilon^{\mu\nu\lambda}\gamma_\lambda .
\end{equation}
 A particular representation is
\begin{eqnarray}\label{eq12}
\gamma^0 &=& I\otimes\sigma_3 ,\;\; \gamma^1 =
i\mbox{diag}(1,-1)\otimes\sigma_2 ,\;\; \gamma^2 =
iI\otimes\sigma_1 .
\end{eqnarray}
We define
\begin{equation}\label{eq40}
   \textbf{a}=-i\gamma^0\gamma^1\gamma^2 =-\gamma^0\sigma^{12}=\mbox{diag}(1,-1)\otimes\sigma_3,
\end{equation}
then the Lagrangian in $2+1$ dimension can further be written as
\begin{eqnarray}\label{eq15}
L =  \bar \psi i \gamma^\mu \partial_\mu \psi - m\bar \psi \psi
 -(1/2){\textbf{a}}\mu_e \epsilon_{\alpha\beta\mu}
\tilde{F}^{\alpha \beta}\bar \psi\gamma^\mu \psi.
\end{eqnarray}

 By using Euler-Lagrange equation, the Dirac equation of motion for a
spin half neutral particle with a electric dipole moment $\mu_e$
is
\begin{equation}\label{eq16}
(i\gamma_\mu
\partial^{\mu}+S_\mu\gamma^\mu-m)\psi =0,
\end{equation}
where
\begin{equation}\label{51}
S_\mu=-(1/2){\textbf{a}}\mu_e \epsilon_{\alpha\beta\mu}
\tilde{F}^{\alpha \beta}.
\end{equation}
The solution to the Dirac equation have the form
\begin{equation}\label{eq17}
\psi=e^{i  \phi_{HMW}}\psi_0,
\end{equation}
where $\psi_0$ is the solution for electromagnetic field free case
. The phase in Eq.(\ref{eq17}) is called HMW phase , and it has
the form
\begin{equation}\label{eq18}
\phi_{HMW}=\int^x S_\mu dx^\mu=-\frac{1}{2}{\textbf{a}}\mu_e
\int^x\varepsilon_{\alpha\beta\mu}\tilde{F}^{\alpha\beta}dx^\mu .
\end{equation}
The HMW phase above is the general HMW phase for a spin-1/2
neutral particle passing through an electromagnetic field. When
the neutral particle moves through a pure static magnetic field,
$\tilde{F}^{\mu\nu}$ reduced to $\tilde{F}^{0i}$, then we have
\begin{equation}\label{eq19}
\phi_{HMW} = -{\textbf{a}}\mu_e
\int^x\varepsilon_{0ij}\tilde{F}^{0i}dx^j = -{\textbf{a}}\mu_e
\int^x(\hat{k}\times \vec{B})\cdot d\vec{x},
\end{equation}
where $\hat{k}$ is the unit vector in $z$ direction and we assume
that the magnetic dipole moment is always along this direction.

\section{ The HMW phase in noncommutative  quantum mechanics}

Now we are in the position to discuss the HMW topological phase in
NC quantum mechanics. First let's consider the NC space case, In
the noncommutative  space the coordinate and momentum operators
satisfy the commutation relations Eq.(\ref{eq1}), and the Dirac
equation for HMW effect is given by Eq.(\ref{eq2}), where $S_\mu$
is given in Eq.(\ref{51}). After the shift defined in
Eq.(\ref{50}), the Dirac equation becomes:
\begin{equation}\label{eq22}
(i\gamma_\mu
\partial^{\mu}-(1/2){\textbf{a}}\mu_e \epsilon_{\mu\alpha\beta} (\tilde{F}^{\alpha\beta}+\frac{1}{2}
 \theta^{\tau\sigma} p_\tau\partial_\sigma\tilde{F}^{\alpha\beta})\gamma^\mu-m)\psi =0.
\end{equation}
This equation is defined in commuting space and the coordinate
non-commutative effect appears in $\theta$ related terms. It is
easy to check that the solution to this Dirac equation has the
form
\begin{equation}
\psi=e^{ i\hat{\phi}_{HMW}}\psi_0,
\end{equation}
where $\psi_0$ is the solution for electromagnetic field free
case, and $\hat{\phi}_{HMW}$ is the HMW phase in NC space, which
is read
\begin{eqnarray}\label{eq23}
\hat{\phi}_{HMW}=-\frac{1}{2}{\textbf{a}}\mu_e
\int^x\varepsilon_{\mu\alpha\beta}\hat{\tilde{F}}^{\alpha\beta}dx^\mu~~~~~~~~~~~~~~~~~~~~~~~~~~~~~~~~~~~~~~~~~~~~~~~~\nonumber\\
 =-\frac{1}{2}{\textbf{a}}\mu_e
\int^x\varepsilon_{\mu\alpha\beta}\tilde{F}^{\alpha\beta}dx^\mu-\frac{1}{4}{\textbf{a}}\mu_e\int^x\epsilon_{\mu\alpha\beta}\theta^{\sigma\tau}
p_\sigma \partial_\tau \tilde{F}^{\alpha
  \beta}dx^\mu.
\end{eqnarray}
This is the general HMW phase for a  spin-1/2 neutral particle
moving in a general electromagnetic field.

    Now let's consider the situation where only static electric field
    exist.
Just like the case discussed in \cite{16}, the Hamiltonian of the
particle in commuting space has the form:
\begin{equation}\label{eq24}
H=\frac{1}{2m}\vec{\sigma}\cdot(\vec{p}+i\mu_e
\vec{B})\vec{\sigma}\cdot(\vec{p}-i\mu_e \vec{B}).
\end{equation}
By using $\vec{\nabla}\cdot \vec{B}=0$,  the Eq.(\ref{eq24}) can
be recast as
\begin{equation}\label{eq25}
H=\frac{1}{2m}(\vec{p}- \vec{\mu}\times\vec{B})^2-\frac{\mu^2
B^2}{2m},
\end{equation}
where $\vec{\mu}=\mu_e\vec{\sigma}$, then the velocity operator
can be gotten
\begin{equation}\label{eq26}
v_l=\frac{\partial H}{\partial
p_l}=\frac{1}{m}[p_l-(\vec{\mu}\times\vec{B})_l].
\end{equation}
From this equation we know that in non-commutative space, we have
\begin{equation}\label{eq27}
p_l=mv_l+(\vec{\mu}\times\vec{B})_l+\mathcal{O}(\theta).
\end{equation}

Insert equation (\ref{eq27}) to (\ref{eq23}) and notice that
\begin{equation}\label{eq28}
\tilde{F}^{\alpha\beta}\longrightarrow \tilde{F}^{0i}~~ {\rm
and}~~
 \theta^{ij}=\theta\epsilon^{ij},~\theta^{0\mu}=\theta^{\mu 0}=0,
\end{equation}
we have
\begin{equation}\label{eq29}
\hat{\phi}_{HMW}=\phi_{HMW}+\delta\phi_{NCS}
\end{equation}
where $\phi_{HMW}$ is the HMW phase in commuting space given by
(\ref{eq19}), the  added phase $\delta \phi_{NCS}$, related to the
non-commutativity of space, is given by
\begin{eqnarray}\label{eq30}
\delta\phi_{NCS}=-\frac{1}{2}{\textbf{a}} \mu_e \int^x
\epsilon_{\mu0i} \theta\epsilon^{\alpha\beta}[mv_\alpha+(
\vec{\mu}\times\vec{B})_\alpha]\partial_\beta \tilde{F}^{0i}dx^\mu ~~~~~~~~\nonumber\\
=\frac{1}{2}{\textbf{a}} \mu_e\theta\epsilon^{ij}\int^x[k_j+(
\vec{\mu}\times\vec{B})_j]
(\partial_iB^2dx^1-\partial_iB^1dx^2),~~
\end{eqnarray}
where $k_j=mv_j$, and the result here coincides with the result
  given in reference \cite{17}, where the tedious star product calculation has been used.

When both space-space and momentum-momentum non-commutating are
considered, i.e. we study the problem on NC phase space, the Dirac
equation for the HMW model is the same as the case on NC space,
but the star product and the shifts are defined in Eqs.(\ref{eq6})
and (\ref{52}). After a similar procedure as in NC space, we got
the Dirac equation on NC phase space as:
\begin{equation}\label{eq33}
\{-\gamma^\mu
p_{\mu}-\frac{1}{2\alpha^2}\gamma^\mu\bar{\theta}_{\mu\nu}x_{\nu}-(1/2){\textbf{a}}\mu_e
\epsilon_{\mu\alpha\beta}[\tilde{F}_{\alpha\beta}+\frac{1}{2\alpha^2}
 \theta^{\tau\sigma} p_\tau\partial_\sigma \tilde{F}_{\alpha\beta}]\gamma^\mu-m' \}\psi
 =0,
\end{equation}
where $m'=m/\alpha$. The solution to (\ref{eq33}) is
\begin{equation}\label{eq34}
\psi=e^{i\hat{\varphi}_{HMW}}\psi_0,
\end{equation}
where $\psi_0$ is the solution of Dirac equation for free particle
with mass $m'$, and $\hat{\varphi}_{HMW}$ stands HMW phase in NC
phase space, and it has the form below,
\begin{eqnarray}\label{eq35}
\hat{\varphi}_{HMW}=-\frac{1}{2}{\textbf{a}}\mu_e
\int^x\varepsilon_{\mu\alpha\beta}\tilde{F}^{\alpha\beta}dx^\mu~~~~~~~~~~~~~~~~~~~~~~~~~~~~~
\nonumber\\-\frac{1}{2\alpha^2 }\int^x \bar{\theta}_{ij}x_jdx_i-
\frac{1}{4\alpha^2}{\textbf{a}}\mu_e\int^x\epsilon_{\mu\alpha\beta}\theta^{\sigma\tau}
p_\sigma \partial_\tau\tilde{ F}^{\alpha
  \beta}dx^\mu.
\end{eqnarray}
Equation (\ref{eq35})is the general HMW phase in noncommutative
phase space. Once again for the case  only static magnetic field
exist, then the HMW phase reduces to
\begin{eqnarray}\label{eq36}
\hat{\varphi}_{HMW}=\phi_{HMW}+\delta\phi_{NCPS},
\end{eqnarray}
where
\begin{eqnarray}\label{53}
\delta\phi_{NCPS}=-\frac{1}{2\alpha^2 }\int^x
\bar{\theta}_{ij}x_jdx_i-\frac{1}{2\alpha^2}{\textbf{a}} \mu_e
\int^x \epsilon_{\mu0i} \theta\epsilon^{\alpha\beta}[m'v_\alpha+(
\vec{\mu}\times\vec{B})_\alpha]\partial_\beta \tilde{F}^{0i}dx^\mu ~~~~~~~~\nonumber\\
=-\frac{1}{2\alpha^2 }\int^x
\bar{\theta}_{ij}x_jdx_i+\frac{1}{2\alpha^2}{\textbf{a}}
\mu_e\theta\epsilon^{ij}\int^x[k'_j+ ( \vec{\mu}\times\vec{B})_j]
(\partial_iB^2dx^1-\partial_iB^1dx^2),~~
\end{eqnarray}
in which $k'_j=m'v_j, ~p_l=m'v_l+
(\vec{\mu}\times\vec{B})_l+\mathcal{O}(\theta)$ has been applied
and we omit the second order terms in $\theta$. The term
$\delta\phi_{NCPS}$ represents the non-commutativity for both
space and momentum. The first term in $\delta\phi_{NCPS}$ is a
contribution purely from the non-commutativity of the momenta, the
second term  is a velocity dependent correction and the third term
is a correction to the vortex of magnetic field. In $2$
dimensional non-commutative plane,
$\bar{\theta}_{ij}=\bar{\theta}\epsilon_{ij}$, and the two NC
parameters $\theta$ and $\bar{\theta}$ are related by
$\bar{\theta}=4\alpha^2 (1-\alpha^2)/\theta$ \cite{6}.  When
$\alpha=1$, which will lead to $\bar{\theta}_{ij}=0$, then the
$\delta\phi_{NCPS}$ returns to $\delta\phi_{NCS}$, namely, the HMW
phase on NC phase space will return to the HMW phase in NC space.

\section{Conclusion remarks }

In this paper, the HMW effect is studied on both noncommutative
space and noncommutative phase space. Instead of doing tedious
star product calculation, we use the "shift" method, i.e. the star
product in Dirac equation can be replaced by Bopp's shift and
together with the shift we defined in (\ref{50}) for NC space and
(\ref{52}) for NC phase space. These shifts is exact equivalent to
the star product. The additional HMW phase terms (\ref{eq23})
(\ref{eq30}) in NC space and the terms (\ref{eq35})(\ref{53}) in
NC phase space are new results of our paper, these two term are
related to non-commutativity of space and phase space. This effect
is expected to be tested at a very high energy level, and the
experimental observation of the effect remains to be further
studied.

The method we use in this paper may also be employed to other
physics problem on NC space and NC phase space. The further study
on the issue will be reported in our forthcoming papers.

\section{Acknowledgments} This paper was completed during our visit
to the high energy section of the Abdus Salam International Centre
for Theoretical Physics (ICTP). We would like to thank Prof. S.
Randjbar-Daemi for his kind invitation and warm hospitality during
our visit at the ICTP. This work is supported in part by the
National Natural Science Foundation of China ( 10575026, and
10447005). The authors also grateful to the support from the Abdus
Salam ICTP, Trieste, Italy.

\end{document}